\documentclass[%
reprint,
 amsmath,amssymb,
 aps,
]{revtex4-2}

\usepackage{graphicx}
\usepackage{dcolumn}
\usepackage{bm}
\usepackage{color}

\begin{document}

\preprint{APS/123-QED}

\title{Polarized quasi-periodic oscillations reveal kink instability in magnetized jets of black holes}

\author{Jiashi Chen}
\author{Pengfu Tian}
\author{Wei Wang}%
 \email{wangwei2017@whu.edu.cn}

\affiliation{Department of Astronomy, School of Physics and Technology, Wuhan University, Wuhan 430072, China}%


%

\date{\today}

\begin{abstract}
The dynamics and instability of the magnetized jets connected to jet acceleration
are complicated and are not yet well understood. Quasi-periodic oscillations (QPOs) as special timing features in black hole systems can directly probe dynamics and structure of accreting and outflow materials. Recently, GHz-band radio polarization oscillations in a stellar-mass black hole are reported, and the physical origin is unclear. We propose that the QPOs in both radio flux and linear polarization will be connected to kink instability in relativistic magnetized jets. The simulations are performed to fit the observed curves of radio flux and linear polarization modulations, in addition, the kink instability model well explains the anti-correlation between flux and linear polarization. These polarized QPOs provide evidence for kink stability in relativistic jets, a phenomenon of significant theoretical importance for understanding the magnetic field configuration near the black hole, as well as for particle acceleration in jets.
\end{abstract}

\maketitle


\textit{Introduction---} Relativistic jets are frequently observed in high-energy astrophysical systems, including active galactic nuclei (AGNs) \cite{2019ARA&A..57..467B} and black hole X-ray binaries (BHXBs) \cite{2004ARA&A..42..317F}. Relativistic jets from accreting black holes (BHs) can radiate non-thermal emission that is highly variable and can lead to flux quasi-periodic oscillations (QPOs) \cite{1999ARA&A..37..409M,2017Natur.552..374R,wang2024cpl}. In recent decades, QPOs have been observed in different mass scales, including supermassive and stellar-mass BHs, and in radio, optical, X-ray and gamma-ray bands with flux modulations \cite{2023Natur.621..271T,2024ApJ...968..106Z,2018NatCo...9.4599Z,2022Natur.609..265J,2024ApJ...974..303Z}. Recently, the GHz-band radio polarization oscillations in a fast-spinning stellar-mass BH GRS 1915+105 were reported by the Five-hundred-meter Aperture Spherical radio Telescope (FAST) \cite{2025arXiv250304011W}: during a radio flare, linear polarization (LP) and flux density (FD) exhibit similar oscillation periods of 17 and 33 seconds with a transient duration of $\sim$ 500 s, and their variation patterns anti-correlate with each other. 

Although QPOs have been observed and studied for decades, their production mechanism is still unclear. Some models are proposed to explain radio QPOs from relativistic jets in BH systems. General relativistic magnetohydrodynamic (MHD) simulations show that a tilted geometrically thin accretion disk can be torn due to the Lense-Thirring torque \cite{2018MNRAS.474L..81L,2021MNRAS.507..983L,2023MNRAS.518.1656M}. QPOs can be observed due to the precession of infalling accretion matter, the disk-jet interaction can be so strong that the jets run into the outer sub-disk \cite{2021MNRAS.507..983L}, leading to instability in jets and QPOs in the radio band. In addition, MHD simulations show that magneto-rotational instability (MRI) can lead to a quasi-periodical shock in the accretion disk and result in a quasi-periodical flux variation \cite{2019PASJ...71...49O,2021RAA....21..134S}, may modulate the magnetic field in jets \cite{2022MNRAS.514.5074O}, thus the MRI model may explain QPOs in both X-ray and radio bands. If A group of recurrent jets with similar periods may be ejected from BH systems, which could also excite the observed quasi-periodic oscillations, and for recurrent jets, the magnetic field may be modulated by some periodic processes, leading to modulations of linear polarization \cite{2022FrASS...932099L}. Finally, a helical motion of emitting blobs in relativistic jets is proposed to explain the quasi-periodic radio-optical light curves of the BL Lac object AO 0235+16 \cite{2004A&A...419..913O}, possible anti-correlations between the degree of polarization and radio flux \cite{2018JApA...39...68M}. Though the recurrent jets model, the MRI shock model, the helical motion of emitting blobs model, and the disk-jets interaction model are potentially able to explain the radio QPOs, however each of them has difficulties in explaining the observed phenomena both in polarized and flux variations in GRS 1915+105 \cite{wang2024cpl}.

Jets are prone to many instabilities, such as pressure-driven (PD), current-driven (CD), and Kelvin-Helmholtz (KH) instabilities (plus any combination of them) \cite{1992A&A...256..354A,2002ApJ...580..800B}. The final nonlinear outcome of these instabilities can reach a simple internal redistribution of jet quantities to the production of internal shocks and/or MHD turbulence \cite{2022A&A...660A..66F}. Jets are highly inhomogeneous media, with tremendous magnetic and velocity gradients whose effects on the development of instabilities are not yet fully understood. The kink instability mechanism is a kind of current-driven plasma instability which causes transverse displacements of plasma and twists the magnetic field structure. It has been suggested that in a rapidly spinning and accreting BH system, theoretical magnetohydrodynamical models predict that a helical magnetic field configuration is naturally produced in the relativistic jets \cite{2008MNRAS.388..551T,2014MNRAS.441.3177M,2021ApJ...906..105C}. When the jet is disturbed at a height with a displacement, the current-driven kink instability will be triggered. The kink can become quasi-periodic, consisting of twisted magnetic field structures, called kink nodes \cite{2017MNRAS.469.4957B,2020MNRAS.494.1817D}. The process results in a moving region (plasma ‘blob’) of enhanced emission, containing a few kink nodes. Due to the quasi-periodic nature of the kink, this blob can exhibit QPO radiation patterns \cite{2021MNRAS.506.1862A}. 

\textcolor{black}{Relativistic magnetohydrodynamic simulations, reported by Dong et al. \cite{2020MNRAS.494.1817D}, have shown that the kinked structure can appear as a periodic signature along the jet direction, which may lead to quasi-periodic oscillation signatures in linear polarization degree and is anti-correlated with the flux light curve \cite{2017ApJ...835..125Z,2020MNRAS.494.1817D}. To calculate the synchrontron emission and polarization signatures, the 3DPol code is used which computes the polarization-dependent synchrotron emission and performs radiation transfer via ray-tracing method \cite{2014ApJ...789...66Z,2018ApJ...862L..25Z,2020MNRAS.494.1817D}. The key physical assumptions in the simulation models are: (i) the jets launches from a simplified central engine of a rotating sphere threaded with magnetic field lines; (ii) the surrounding interstellar gas follows a broken power-law density profile; (iii) the initial jet plasma and the surrounding medium are cold and laminar; (iv) the non-thermal particles follow a power-law distribution whose energy density is proportional to the local thermal energy density. The kink instability is suggested in blazer jets and has been used to explain QPO features observed in some supermassive BH systems \cite{2022Natur.609..265J,2024MNRAS.527.9132T}. The key assumptions can be achieved in stellar black hole systems, so that the simulations of kink can be applied in the BH jets in different scales.}

\begin{figure*}
    \centering
    \includegraphics[width=\textwidth]{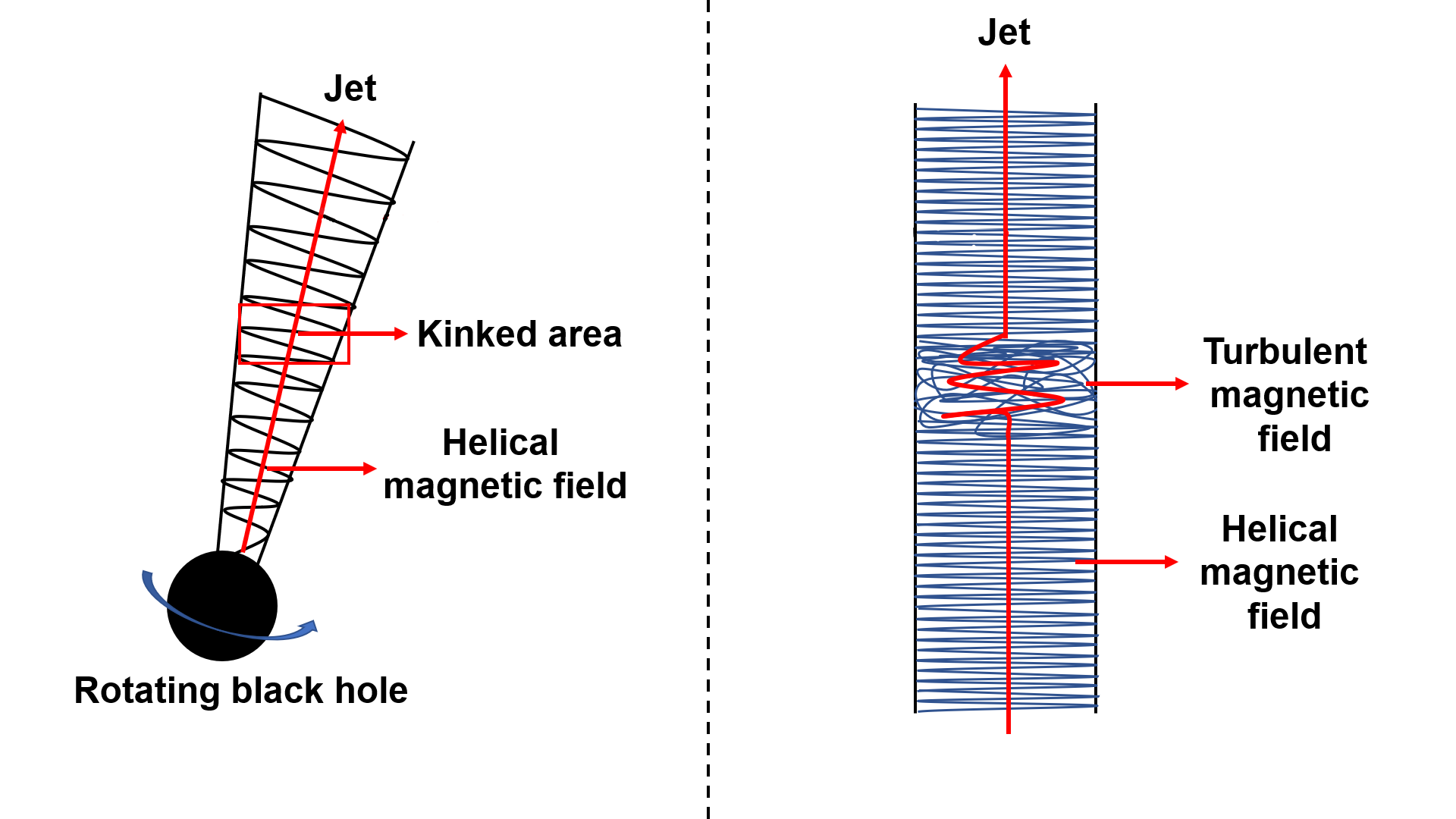}
    \caption{Sketch of kink instability of the magnetized relativistic jet in a black hole system. {\bf Left panel:} The helical magnetic field structure of the relativistic jet above the rotating black hole system is expected. In the region far away from the BH, the instability would develop in the kinked area of the magnetized jet. {\bf Right panel:} Kink instability will cause transverse displacements of plasma and twist the magnetic field structure, then dissipate a significant amount of magnetic energy and accelerate non-thermal particles. Kink instabilities cause the quasi-periodic magnetic energy conversion to thermal energy, which leads to a quasi-periodic emission signature.}
    \label{kink1}
\end{figure*}

The kink instability has been evidenced in solar flares and can be connected to a magnetic mechanism for solar eruptive phenomena \cite{2005ApJ...630L..97T,2006PhRvL..96y5002K}. In addition, the observed optical flux QPOs at $\sim 0.6$ day in BL Lacertae during its 2020 outburst \cite{2022Natur.609..265J} were explained with the kink instability in the jets from the supermassive BH. Optical flux quasi-periodic oscillations of two other supermassive BH systems with periods around 2-6 days, 1RXS J111741.0+254858 and 1RXS J004519.6+212735, may be due to the kink instability in relativistic jets \cite{2024MNRAS.527.9132T}. However, in these supermassive BH systems, only the flux light curves show the possible QPO features; there is no evidence for the polarization oscillations or lack of polarization observations. Thus, in accreting BH systems, we still lack evidence for kink stability in jets. In this Letter, we perform numerical simulations of the light curves of radio flux density and polarization parameters considering kink instability in magnetized jets, and compare with the FAST data of GRS 1915+105. This work provides evidence that kink instability in jets leads to polarized QPOs.


\textit{Method---} The kink instability can develop in a magnetized relativistic jet with a helical magnetic field configuration \cite{2009ApJ...700..684M,2020MNRAS.494.1817D} (a brief sketch is shown in Figure \ref{kink1}). The growth of the kink in the jet will develop in the jet region where the magnetic fields have opposite directions and reconnect to accelerate the electrons. Both the poloidal field component and the number of accelerated particles increase at the same location \cite{2017ApJ...835..125Z,2021MNRAS.501.2836B}. The kink instability can dissipate significant amounts of magnetic energy to accelerate particles \cite{2009ApJ...700..684M}. Depending on the setup of the simulations, about 10-50\% of magnetic energy can be dissipated, which will be transferred to accelerated particles with a power-law spectrum \cite{2018PhRvL.121x5101A,2020ApJ...896L..31D}. The kink instability can distort the magnetic field and generate an induced electric field. The combination of induced electric field and distorted magnetic field is an effective non-thermal particle accelerator. Particles accelerated through kink instability can give rise to synchrotron emission, which accounts for observed polarization signatures \cite{2017ApJ...835..125Z}.

For the black hole system GRS 1915+105, both the radio flux and linear polarization observed by FAST show similar quasi-periodic oscillation signatures with the same period of about 17 and 33 seconds \cite{2025arXiv250304011W}. In addition, the variation patterns of radio flux and linear polarization show anti-correlation behaviors. Therefore, we argue that the observed polarized QPOs may be generated by kink instability in the BH jets \textcolor{black}{, and we adopted the kink model built by L. Dong and H. Zhang \cite{2020MNRAS.494.1817D} (also see Method section in S. G. Jorstad et al. \cite{2022Natur.609..265J}, which has used this method to explain the observed quasi-periodic oscillations in the relativistic jet of BL Lacertae, and the codes can be found in the supplemental materials of this reference) to simulate kink process. In the followings, we used the Markov Chain Monte Carlo (MCMC) method based on the above kink instability model to simulate kink instability process and parameters in the relativistic magnetized jets and fit the radio data of GRS 1915+105 observed by FAST with simulation results.}

In this \textcolor{black}{model}, it assumes that highly magnetic jets launch from a spinning black hole, and the accelerated electrons follow a power-law distribution. The spin of the black hole will generate a helical magnetic field, and the kink instability will naturally develop in this system \cite{2009ApJ...700..684M,2017MNRAS.469.4957B}. The \textcolor{black}{model} assumes a simple magnetic field structure: a constant toroidal component whose flux is normalized to unity in the code, a periodically fluctuating poloidal contribution characterized by a sinusoidal function with a period $T$ and an amplitude $B_{p0}$, a turbulence whose contribution averages to $P_0$ and with an amplitude of fluctuations of $B_0$. The average emission power in the code unit is normalized to $F_{r0}$ for the radio bands. The \textcolor{black}{model} consists of two steps: a MCMC fitting to constrain the above parameters, and then calculating the model with the 50\% quantile (median) of the fitted values of each parameter \textcolor{black}{according to the following equations:}

\begin{align}
    B_{p}(t) &=B_{p0}*sin(\frac{t}{T}*2\pi+\varphi) \\
    B_{tur}(t)&=B_{0}*sin(\frac{t}{T}*2\pi+\varphi) \\
    EM_{ave}&=B_{tor}^2+0.5*B_{p0}^2+P_0 \\
    EM_{temp}(t)&=B_{tor}^2+B_{p}(t)^2+B_{tur}(t)^2 \\
    factor&=\delta^{\alpha+3} \\
    FD(t)&=F_{r0}*\frac{EM_{temp}}{EM_{ave}}*factor \\
    LP(t)&=\frac{\vert {B_{tor}^2-B_{p}(t)^2} \vert}{EM_{temp}}  * C,
\end{align}
where $B_{tor}$ is the toroidal magnetic field component and is set to unity in the code, $B_{p}(t)$ is the poloidal magnetic field component, the $B_{tur}(t)$ is the fluctuation of the magnetic field, $EM_{ave}$ is the average magnetic energy density and $EM_{temp}(t)$ is the magnetic energy density at time $t$, $\delta$ is the Doppler factor, we set the constant $\alpha$ to 1.2 and $C$ to 77 according to the simulations of Dong et al. \cite{2020MNRAS.494.1817D}.

\begin{figure*}
    \centering
    \includegraphics[width=0.8\textwidth]{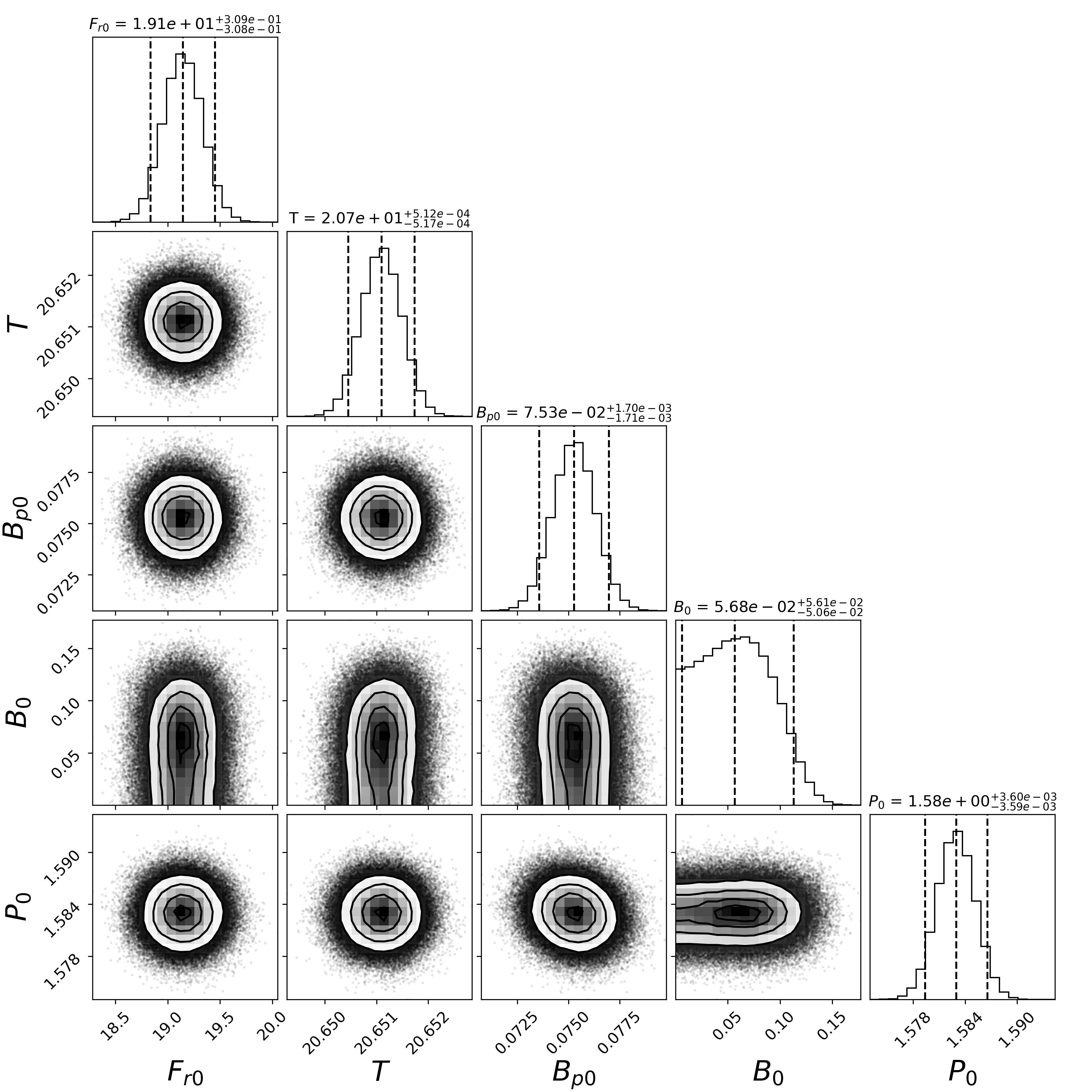}
    \caption{Triangle plots of posterior distributions of model parameters for kink instability in the relativistic jet of GRS 1915+105.}
    \label{figure1}
\end{figure*}

\textit{Result---} Fig. \ref{figure1} shows the MCMC results for each parameter in kink stability. We find that the full period of the twist in the kink is about 20.7 seconds, which is slightly larger than the QPO period (17 s) obtained from the average power spectrum of the flux density. Such a difference is also seen in BL Lacertae, where the period obtained from the MCMC is slighter larger than twice of the observed QPO in optical light curve. The difference in period is probably due to the turbulence component in the model, which is estimated to be relatively strong in the MCMC simulation. 

The QPO period is believed to be associated with the growth timescale of kink structures, which can be estimated by the evolution of the transverse motion \cite{2009ApJ...700..684M}. \textcolor{black}{This timescale can be estimated by the ratio of the transverse displacement of the jet from its central spine to the average transverse velocity, $\tau_{KI}\sim R_{KI}/\langle v\rangle_{tr}$, where $R_{KI}$ is the transverse displacement of the strongest kinked region and $\langle v\rangle_{tr}\sim 0.1 c$ is the average transverse velocity, which can be estimated in a detailed relativistic MHD simulation \cite{2020MNRAS.494.1817D}, and $c$ is the light speed. Thus a simulated period of 20.7 s corresponds to a transverse displacement size of the jet of $R_K\approx 10^{11}$ cm. FAST observations suggest that the rise timescale of the radio flux from the start to the trigger of kink instability is $t\sim 1000$ s \cite{2025arXiv250304011W}, whereupon the jet height of the kink region $h\sim ct\approx 3\times 10^{13}$ cm. A ratio $R_K/h \lesssim 1\%$ suggests that the relativistic jet is highly collimated, consistent with predictions of numerical simulations of the jet spine collimated by strong magnetic fields twisted in the rotating BH \cite{2009ApJ...700..684M}.}

\textcolor{black}{The MCMC results give} a relatively strong turbulence, and the amplitude is close to the amplitude of the periodically fluctuating poloidal component. In the non-turbulent case, there is a strong anti-correlation between flux and polarization (with the Pearson correlation coefficient of $\sim-1$) \cite{2020MNRAS.494.1817D}. When turbulence is considered, the anti-correlation may become statistically insignificant. In GRS 1915+105, when the flux and linear polarization have similar QPO periods, an anti-correlated relationship, with the Pearson correlation coefficient of -0.795 and a slope of -0.04, is observed between flux and linear polarization \cite{2025arXiv250304011W}. The Pearson correlation coefficient is around 0.2 larger than the non-turbulent case, which is due to the effect of turbulence.

\begin{figure*}
    \centering
    \includegraphics[width=\textwidth]{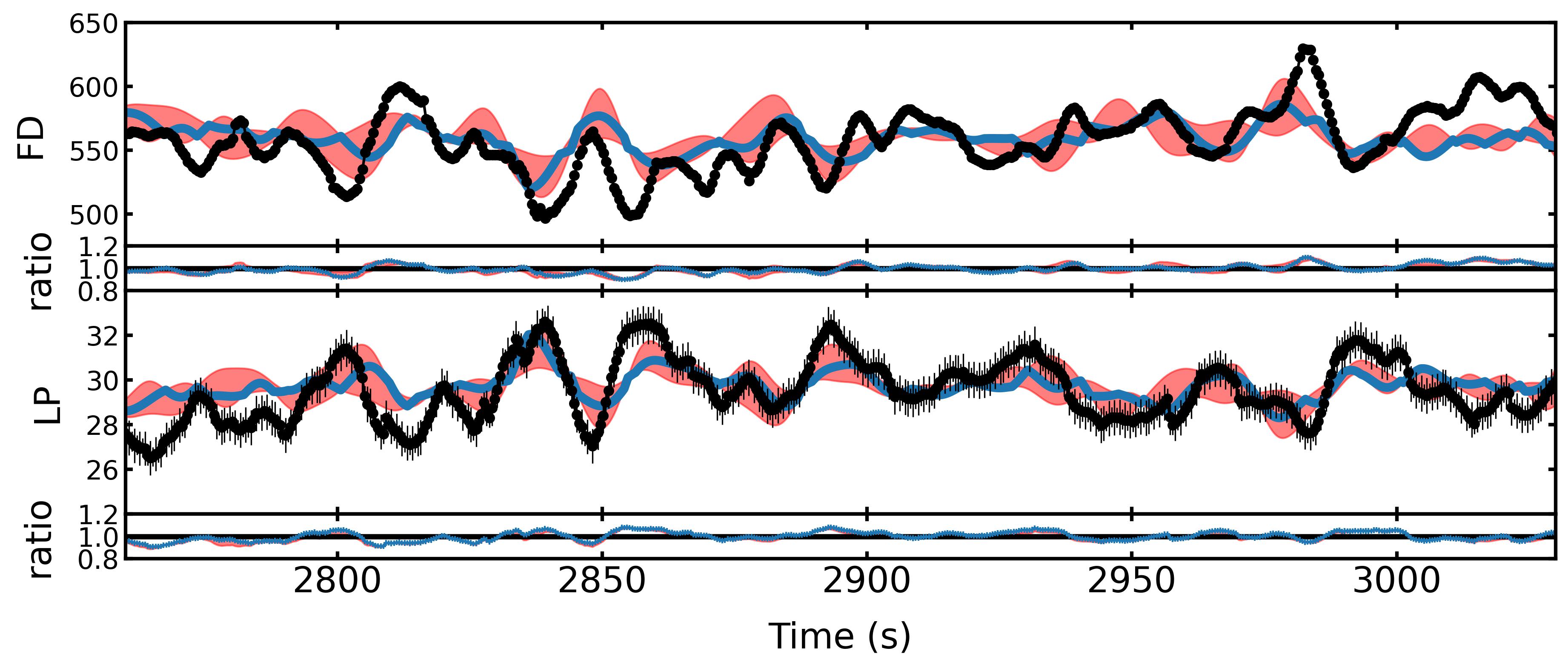}
    \caption{Comparison between the FAST data \cite{2025arXiv250304011W} and kink instability model with the ratio of data to model is plotted below. Black symbols correspond to the observed data of flux density (FD), linear polarization (LP), and the blue curves show the fit of a kink instability model using an MCMC code. The red shaded region is calculated using up and low limits of the amplitude of fluctuations for $B_0$.}
    \label{figure2}
\end{figure*}

We then calculated the model with the 50\% quantile (medium) of the fitted values of each parameter. Fig. \ref{figure2} shows the model-to-data comparison plot, where the black dots are observed data, the solid blue line is the simulation result of the kink instability model and the ratio of data to model is plotted below. The red shaded region is calculated using up and low limits of the amplitude of fluctuations for $B_0$. The model can broadly fit the peaks and dips of the flux and linear polarization curves between 2800-3000 s, when strong QPO signatures are observed (see Fig. 3 in \cite{2025arXiv250304011W}).

It has been suggested that kink instability is connected to the azimuthal twist of magnetic tubes, trigger solar flares and coronal mass ejections \cite{2010ApJ...715..292S}. Observations of kink instability accompanied with full filament eruption \cite{2005ApJ...628L.163W}, partial cavity eruption \cite{2007ApJ...661.1260L}, partial filament eruption \cite{2008ApJ...680.1508L}, and failed filament eruption \cite{2006ApJ...653..719A} indicate its importance in filament interaction with magnetic mechanism for solar eruptive phenomena \cite{2005ApJ...630L..97T,2006PhRvL..96y5002K}. The kink instabilities in eruptions can lead to oscillations in the flux \cite{2012ApJ...746...67K,2014A&A...572A..83K}, which is similar to our observations of GRS 1915+105. The kink instabilities could be the key to understand the efficient non-thermal ion and electron acceleration \cite{2021PhRvL.127r5101Z}, the origin of $\delta$-spot regions \cite{2015ApJ...813..112T}, magnetic twist \cite{2018ApJ...853...35T} and structure of magnetic fields \cite{2016ApJ...832..106H} in solar flares. The present work also shows that the polarized oscillations provide evidence for the kink instability occurring in magnetized relativistic jets of a BH system.

\textit{Conclusion---}In this paper, we have studied the QPO production in radio flux and linear polarization that are observed in GRS 1915+105. We show that kink instability in relativistic jets can generate these observed QPOs when the magnetic field around the BH consists of a constant toroidal component, a periodically fluctuating poloidal contribution characterized by a sinusoidal function with a period around 20.7 s. The anti-correlation between radio flux and linear polarization can also be explained by the kink instability model. These make the kink instability a promising origin of the radio QPOs in GRS 1915+105 and particle acceleration in magnetized jets.

\nocite{*}

\bibliography{kink.bbl}

\end{document}